\documentclass[final,3p,times,twocolumn]{elsarticle}

\usepackage{amssymb}
\usepackage[dvipsnames]{xcolor}
\usepackage{amsmath}
\usepackage{soul}
\usepackage{comment}
\journal{Physics Letters B}

\begin{document}

\begin{frontmatter}

\title{Gravitational form factors of the pion and meson dominance}

\author[1,2]{Wojciech Broniowski}
\ead{Wojciech.Broniowski@ifj.edu.pl}

\author[3]{Enrique Ruiz Arriola}
\ead{earriola@ugr.es}

\affiliation[1]{organization={H. Niewodniczanski Institute of Nuclear Physics PAN},
postcode={31-342},
city={Cracow},
country={Poland}}

\affiliation[2]{organization={Institute of Physics},
addressline={Jan Kochanowski University},
postcode={25-406},
city={Kielce},
country={Poland}}

\affiliation[3]
%{organization={Departamento de F\'{\i}sica At\'{o}mica, Molecular y Nuclear and Instituto Carlos I de  F{\'\i}sica Te\'orica y Computacional},
{organization={Departamento de Fisica Atomica, Molecular y Nuclear and Instituto Carlos I de Fisica Teorica y Computacional}, 
addressline={Universidad de Granada},
postcode={E-18071},
city={Granada},
country={Spain}}

\begin{abstract} 
We show that the recent MIT lattice QCD data for the pion's gravitational form factors are,  in the covered momentum transfer range, 
fully consistent with the meson dominance principle. In particular, the $2^{++}$ component can be 
accurately saturated with the $f_2(1270)$ meson, whereas the 
$0^{++}$ component with the $\sigma$ meson. 
To incorporate the large width of the $\sigma$, we use the dispersion relation with the spectral density obtained 
from analyses of the physical pion scattering data. Effects of the pion mass are estimated within Chiral Perturbation Theory and are found to be small 
between the lattice and the physical point. 
We also discuss the implications of the perturbative QCD constraints at high momentum transfers, leading to specific sum rules
for the spectral densities of the gravitational form factors, and argue that these densities cannot be of definite sign.
\end{abstract}


\end{frontmatter}

% \linenumbers

\bibliographystyle{elsarticle-num-names} 

{\bf Introduction.} Recently, high precision lattice QCD data for the gravitational form
factors (GFFs) of the pion were
released~\cite{Hackett:2023nkr,Pefkou:2023okb}, with the pion mass
$m_\pi=170$~MeV close to the physical point, and including all the
species of partons (the light quarks and gluons).
%\footnote{This unphysical pion mass corresponds to a relative change of $ s - 4 m_\pi^2 \sim
%  \tilde s -4 \tilde m_\pi^2 $ which corresponds to a shift of $ s = 4 \sim 4(\tilde m_\pi^2 - m_\pi^2 ) =0.03 {\rm GeV^2} $.}.  
This vastly improves on the early simulations of the
quark contributions~\cite{Brommel:2007zz,QCDSF:2007ifr}, recently
repeated in~\cite{Delmar:2024vxn} for the pion masses $\sim 250$~MeV,
or for the gluonic contributions~\cite{Shanahan:2018pib} and the
gluonic scalar component (trace anomaly)~\cite{Wang:2024lrm}, studied
at large pion masses.  The accuracy of the data
of~\cite{Hackett:2023nkr,Pefkou:2023okb} and the proximity to the
physical limit allows for more stringent comparisons with theoretical
expectations and models.  GFFs, describing the mechanical properties
of hadrons, have been discussed since the work of Pagels dating back
to 1965~\cite{PhysRev.144.1250} (for a review and literature see,
e.g.,~\cite{Polyakov:2018zvc}).  Numerous model calculations for the pion have been carried out, see in
particular~\cite{Broniowski:2007si,Broniowski:2008hx,Frederico:2009fk,Masjuan:2012sk,Fanelli:2016aqc,Freese:2019bhb,Krutov:2020ewr,Xing:2022mvk,Xu:2023izo,Li:2023izn}, recently also in the instanton liquid model~\cite{Liu:2024jno,Liu:2024vkj}.
Importantly, an extraction of the pion GFFs has been
inferred from the $\gamma \gamma^\ast \to \pi^0 \pi^0$ experimental
data%
%~\cite{Belle:2015oin} in
~\cite{Kumano:2017lhr}, using the link of
GFFs to the generalized distribution amplitudes.

% (* Perhaps link and references to the dark matter searches *)

{\bf Definitions.} The GFFs of the pion correspond to the matrix elements of the stress-energy-momentum (SEM) tensor $\Theta^{\mu\nu}$
between on-shell pion states,
\begin{eqnarray}
&& \hspace{-7mm} \langle \pi (p') | \Theta^{\mu \nu}(0) | \pi (p) \rangle \equiv \Theta^{\mu \nu} =  \nonumber \\
&& 2 P^\mu P^\nu A(q^2)  + \frac12 ( q^\mu q^\nu - g^{\mu \nu} q^2 ) D(q^2 ),
\end{eqnarray}
where $P=\frac12 (p+p')$, $q=p'-p$, and $t=q^2$. On-shell, one has $P^2 = m_\pi^2 - q^2/4$ and $P \cdot q=0$. 
We omit for brevity the isospin indices of the pion, as the considered operator is isoscalar.
We consider the full SEM operator, summing up the contributions from all the quarks species and gluons, 
$\Theta^{\mu \nu}=\sum_p \Theta^{\mu \nu}_p$, whose matrix elements are conserved, $q_\mu \Theta^{\mu \nu}(q^2)=0$, as well as 
renormalization scale and scheme independent~\cite{Hackett:2023nkr,Ji:1994av,Lorce:2017xzd,Hatta:2018sqd}.  
The trace is given by 
\begin{eqnarray}
 \Theta^{\mu}_\mu(t) \equiv \Theta(t) = 2 \left(m_\pi^2- \tfrac{1}{4} t \right) A(t) - \tfrac{3}{2} t\, D(t). \label{eq:th0}
\end{eqnarray}

The Lorentz invariant vertex functions $A(t)$ and $D(t)$, as well as $\Theta(t)$, obey a number of constraints based on relativity,
analyticity, unitarity, chiral symmetry, and pQCD, which provide
a  basic qualitative understanding of their $t$ dependence, as we shall discuss later.
In particular, from the mass sum rule $\langle \pi (p) | \Theta^{\mu \nu}(0) | \pi (p) \rangle =  2 p^\mu p^\nu$ one infers 
the normalization condition $A(0)=1$, hence $\Theta(0)=2m_\pi^2$,
whereas a chiral Ward identity yields a low energy theorem in
Chiral Perturbation Theory ($\chi$PT)~\cite{Novikov:1980fa,Donoghue:1991qv},
\begin{eqnarray}
D(0)=-1+{\cal O}(m_\pi^2), \;\;\; \Theta(t)=2m_\pi^2+t+{\cal O}(t^2, t m_\pi^2). \label{eq:cons}
\end{eqnarray} 

The rank-two tensor $\Theta^{\mu \nu}$ can be decomposed into a sum of two separately conserved irreducible tensors corresponding to a
well-defined total angular momentum,  $J^{PC}=0^{++}$ (scalar) and $2^{++}$ (tensor), namely~\cite{Raman:1971jg}
\begin{eqnarray}
&& \Theta^{\mu \nu} =  \Theta_S^{\mu \nu}+\Theta_T^{\mu \nu},  \label{eq:thetaS} \\
&&  \Theta_S^{\mu \nu} = \frac13 Q^{\mu \nu} \Theta, \nonumber \\ 
&&  \Theta_T^{\mu \nu} = \Theta^{\mu \nu}- \frac13  Q^{\mu \nu} \Theta  = 2 \left[ P^\mu P^\nu - \frac{P^2}3  Q^{\mu \nu} \right] A, \nonumber
\end{eqnarray} 
where $Q^{\mu \nu}\equiv g^{\mu \nu}-{q^\mu q^\nu}/{q^2}$.
Since $\Theta$ and $A$ carry the information on good $J^{PC}$ channels, they should be regarded as the primary objects, whereas the $D$
form factor mixes the quantum numbers, and reads 
\begin{eqnarray}
D= -\frac{2}{3t} \left [ \Theta - \left ( 2 m_\pi^2 -\tfrac{1}{2}\, t \right ) A \right]. \label{eq:Drel}
\end{eqnarray}

\begin{figure}[tb]
\begin{center}
\includegraphics[angle=0,width=0.4 \textwidth]{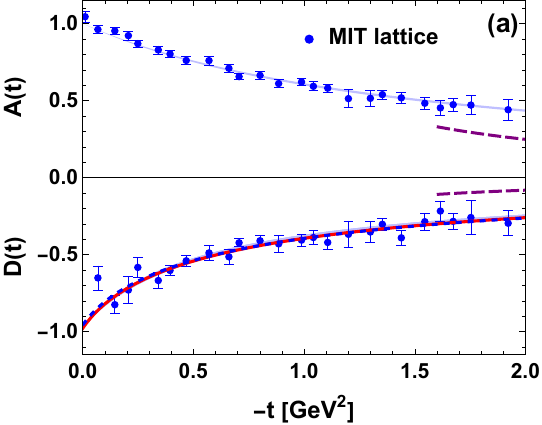} \\
\includegraphics[angle=0,width=0.4 \textwidth]{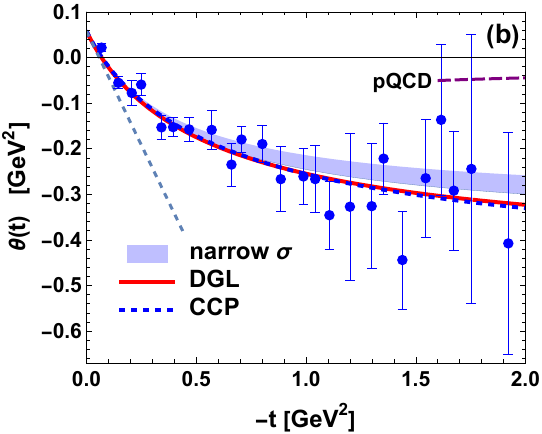} 
\end{center}
\vspace{-5mm}
\caption{Gravitational form factors of the pion, plotted as functions of the space-like momentum transfer $-t=Q^2$: (a) Form factors $A(t)$
and $D(t)$ (thin bands, with widths indicating the uncertainty of the fit), compared to the data from~\cite{Hackett:2023nkr} (points with error bars). For $A$, the monopole formula~(\ref{eq:mf}) with $m_{f_2}=1.24(3)$~GeV is used, while $D$ is obtained
from Eqs.~(\ref{eq:Drel},\ref{eq:mf},\ref{eq:ms}). (b) The  trace anomaly form factor $\Theta(t)$. The data are obtained
from~\cite{Hackett:2023nkr} via Eq.~(\ref{eq:Drel}), with errors added in quadrature. The tangent (dashed line) at the origin corresponds to the
$\chi$PT formula $\theta(t)\simeq 2m_\pi^2+t$. The band shows the narrow resonance approximation with $m_{f_2}=1.24(3)$~GeV and
$m_{\sigma}=0.65(3)$~GeV. The model curves DGL and CCP are obtained for the spectral densities shown in Fig.~\ref{fig:spect},
used in the dispersion relation~(\ref{eq:Th}).  The long-dashed
lines in both panels are the asymptotic pQCD results~(\ref{eq:AD-asymp},\ref{eq:Theta-asymp})~\cite{Tong:2022zax} for $\Lambda_{\rm QCD}= 225~{\rm MeV}$;
the range of the lattice QCD data is far from reaching asymptotics.
The data and the narrow resonance fit correspond to $m_\pi=170$~MeV, whereas DGL and CCP are at the physical point  $m_\pi=140$~MeV.  \label{fig:fit} }
\end{figure}

{\bf Asymptotics.} The leading-order  perturbative QCD (pQCD) asymptotics at $t \to -\infty$ has been determined in~\cite{Tong:2021ctu,Tong:2022zax} 
(recently corroborated in~\cite{Liu:2024vkj}), yielding\footnote{This can be readily
  obtained from Eqs.~(5,6) in~\cite{Tong:2021ctu} by using the
  asymptotic light cone pion wave function $\phi(x) = \sqrt{6} f_\pi x(1-x)$.}
\begin{eqnarray}
  A(t)=-3 D(t)  \left( 1\! +\! {\cal O} (\alpha) \right) = -\frac{48 \pi \alpha(t) f_\pi^2} {t} \left( 1 \!+\! {\cal O} (\alpha) \right), 
  \label{eq:AD-asymp}
\end{eqnarray}  
where $\alpha(t) = (4\pi /\beta_0) / \ln \left (-t/\Lambda_{\rm QCD}^2 \right) $ is the running strong coupling
constant with  $ \beta_0 = \tfrac{1}{3} (11 N_c- 2 N_f)$ with $N_c =3$ colors and $N_f$ active flavors, whereas $f_\pi$
% ** = 93$~MeV
denotes the pion weak decay constant. Thus $A$ and $D$ approach 0 as $1/Q^2$ (up to logarithmic corrections)
from the positive and negative sides, correspondingly (cf. the long-dash lines in Fig.~\ref{fig:fit}a).
%To analyze the high energy limit, we recall that the
The trace anomaly of QCD reads
\begin{eqnarray}
\Theta(t) =\frac{\beta(\alpha)}{4\alpha} {G^{\mu \nu}}^2+ [1+\gamma_m(\alpha)] \sum_f m_f \bar{q}_f q_f, \label{eq:anom}
\end{eqnarray} 
where $\beta(\alpha) = \mu \, d \alpha / d \mu = - \alpha [\beta_0 (\alpha/2\pi) +
%\beta_1 (\alpha/4\pi)^2 +
{\cal O}(\alpha^2)  ] < 0 $ is the QCD beta function, $\gamma_m(\alpha)= 2 \alpha/\pi +{\cal O}(\alpha^2) $ is the quark mass anomalous dimension,
 and  $f$ enumerates the active flavors.
%
% At the physical pion mass, the quark contribution yields about 25\%
% to the normalization at $Q^2=0$~\cite{Gasser:1982ap,Ji:1995sv}. 
From pQCD,
the leading twist asymptotic behavior of the scalar-isoscalar form factor related to the chirally odd quark component of~(\ref{eq:anom}) can be written 
as $m_q \langle \pi (p') | \bar{q} q (0) | \pi(p) \rangle \sim \, m_\pi^2 f_\pi^2 \alpha(t)/t$~\cite{Lepage:1980fj}, which approaches 0 as $1/Q^2$ up to logarithmic corrections. 
The situation is different, however, for the 
gluonic part, where at $Q^2 \to \infty$~\cite{Tong:2022zax,Liu:2024vkj} % the absolute contributions becomes 
\begin{eqnarray}
  && \hspace{-7mm}\langle \pi (p') | \frac{\beta(\alpha)}{4 \alpha} {G^{\mu \nu}}^2(0) | \pi(p) \rangle  = 8 \pi \beta \left (\alpha(t) \right) f_\pi^2 + {\cal} O(\alpha^3) \nonumber \\ && = -4 \beta_0 \alpha(t)^2 f_\pi^2 + {\cal} O(\alpha^3),
  \label{eq:Theta-asymp}
\end{eqnarray}
which means that $\Theta(-Q^2)$ goes to zero from the negative side very slowly, as a negative constant 
divided by $( \ln Q^2/\Lambda_{\rm QCD}^2)^2$ (see the long-dash line in Fig.~\ref{fig:fit}b).
This flatness is also quite spectacularly seen in the lattice simulations of the gluonic trace anomaly~\cite{Wang:2024lrm}, which 
extend to $Q^2\sim 3.7~{\rm GeV}^2$ (albeit with pion masses significantly higher from the physical value). Note that the leading
behavior in $\alpha(t)$ of Eq.~(\ref{eq:AD-asymp}) cancels in Eq.~(\ref{eq:Theta-asymp}) due to relation (\ref{eq:th0}).

%(* Interesting conclusions for the spectral density may be drawn from the recently obtained asymptotic behavior of 
%the components of $\Theta(Q^2)$. *)  
  
{\bf Dispersion relations.} Next, we review analyticity features of GFFs.
The functions $A(t)$, $D(t)$, and $\Theta(t)$ 
are real in the space-like region ($t < 0$), and 
develop branch cuts at the
$2\pi, 4\pi, K \bar K$, etc. production thresholds, corresponding to
$t= 4 m_\pi^2, 16 m_\pi^2, 4 m_K^2$, etc.
From analyticity and with the above-discussed limits at small and large
momenta, GFFs satisfy dispersion relations, which in a once-subtracted
form are
\begin{eqnarray}
  A(t) &=&      1+\frac1{\pi} \int_{4 m_\pi^2}^\infty ds \frac{t}{s}\frac{{\rm Im} \,A(s) }{s-t-i \epsilon}, \label{eq:A}\\
  D(t) &=& D(0)+\frac1{\pi} \int_{4 m_\pi^2}^\infty ds \frac{t}{s}\frac{{\rm Im} \,D(s) }{s-t-i \epsilon}. \label{eq:D}
\end{eqnarray}
Here $ 2 i\, {\rm Im} \,F(s) \equiv  {\rm disc} \,F(s) \equiv F(s\!+\! i\epsilon)-F(s\!-\!i \epsilon)$ denotes the usual
discontinuity,  with the spectral density ${\rm Im}\, F(s)/\pi$ along the branch cut. %at $s > 4 m_\pi^2$. 
Similarly,%since $\Theta (t) \sim \alpha^2 f_\pi^2$
%~\cite{Tong:2021ctu,Tong:2022zax}~\footnote{Note the cancellation to
%  ${\cal \alpha}$ from Eq.~(\ref{eq:AD-asymp}), connected to scale
%  invariance and its anomalous breaking, see also below} one has
\begin{eqnarray}
  \Theta(t)  &=& 2 m_\pi^2 + \frac1{\pi} \int_{4 m_\pi^2}^\infty ds \frac{t}{s}\frac{{\rm Im} \, \Theta (s)}{s-t-i\epsilon}. \label{eq:Th}
\end{eqnarray}

%With the above dispersion relations, 
Obtaining and explaining the
behavior of GFFs in the space-like region $-t=Q^2\ge 0$ may be viewed
as modeling the discontinuities along the cut, which corresponds to
analyzing the $s$-channel matrix element $ \langle \pi \pi |
\Theta^{\mu \nu} | 0 \rangle $.  One typically distinguishes three
domains: the region $4 m_\pi^2 \le s \le \Lambda_\chi^2$ close to the
production threshold, where $\chi$PT sets in, the region of large values $s >
\Lambda_{\rm pQCD}^2$  where pQCD can be
used, and the remaining intermediate region, $s\sim 1-3 {\rm ~GeV}^2$,
where meson resonances are dominant. Numerical values for
$\Lambda_\chi $ and $\Lambda_{\rm pQCD}$ will be discussed
below. With the explanation of the recent lattice QCD results as a
primary goal, we begin with the resonance region, discussing the {\em a posteriori} small $\chi$PT and pQCD effects afterwards.

{\bf Narrow resonances.} The intermediate energy region can in principle be handled by means of
final state interactions, using for unitarization  the
Omn\`es~\cite{Donoghue:1990xh} or Bethe-Salpeter~\cite{Nieves:1999bx}
representations applied to the matrix element $\langle \pi \pi |\Theta^{\mu \nu} | 0 \rangle$, hence implementing Watson's final
state theorem for the $\pi\pi $ scattering,
$\Theta(s+i\epsilon)=e^{2 i \delta_{\pi\pi}(s)} \Theta (s-i \epsilon)$,
in the elastic region $4 m_\pi^2 \le s \le 4 m_K^2$. 
%(see below). 
Nonetheless, the impact of unitarization in the time-like
region onto the space-like region is mild and resonating phase-shifts
can be effectively replaced by a step function mimicking a
monopole with the mass possibly shifted from the nominal Breit-Wigner (BW) value
$\delta(m_R^2)=\pi/2$~\cite{Masjuan:2012sk}.\footnote{This can be seen from the Omn\`es representation and works even for a broad scalar resonance,
% Whereas resonances are properly defined as poles in the second Riemann sheet of the
% complex plane $s_R= m_R^2 - i \Gamma_R m_R $, the previous statement does not apply to the pole position directly, but rather to the BW
%  definition for {\it wide} resonances, 
since $\delta(s_{BW})=\pi/2 + {\cal O} (N_c^{-3})$~\cite{Nieves:2009kh}.} 
We thus content ourselves for the moment with the narrow resonance
approximation~\cite{Ecker:1989yg,Ecker:1988te} which befits the
large-$N_c$ limit~\cite{Pich:2002xy,Ledwig:2014cla} and explicitly
realizes the tensor decomposition of Eq.~(\ref{eq:thetaS}). 

Applying the standard resonance saturation by inserting a complete set of intermediate hadronic states yields
\begin{eqnarray}
  \langle \pi \pi  | \Theta^{\mu \nu} | 0 \rangle = 
 \sum_R \langle \pi \pi | R \rangle \frac{1}{m_R^2-q^2} \langle R | \Theta^{\mu \nu} | 0 \rangle,
\end{eqnarray}
with the contributions limited to $0^{++}$ and $2^{++}$ states.
The field representation of higher spin particles, such as the tensor
mesons, is not unique when particles are not on-shell and
generically produces polynomial pieces diverging at large
energies.\footnote{An
  example is provided by the contribution of the $f_2$ exchange to the
  $\pi\pi$ scattering in the large-$N_c$ limit, whose dispersive
  pieces violate the Froissart
  bound~\cite{Toublan:1995bk,Ecker:2007us,Nieves:2011gb}. } Hence, it is far more practical to compute the absorptive
part of the form factor, which at $q^2 \to s+ i \epsilon$ reads
\begin{eqnarray}
\frac1{\pi} {\rm Im}  \langle \pi \pi  | \Theta^{\mu \nu} | 0 \rangle = 
 \sum_R \langle \pi \pi | R \rangle  \langle R | \Theta^{\mu \nu} | 0 \rangle \delta(m_R^2-s), 
\end{eqnarray}
and then reconstruct the dispersive part from the dispersion relation with suitable subtraction constants. The vacuum
to hadron transition amplitudes are
\begin{eqnarray}
  \langle S | \Theta^{\mu \nu} | 0 \rangle = \tfrac{1}{3} f_S  q^2 Q^{\mu\nu}, \;\;\; \langle T | \Theta^{\mu \nu} | 0 \rangle = f_T m_T^2 \epsilon^{\mu \nu}_\lambda, 
\end{eqnarray}
where $\epsilon^{\mu \nu}_\lambda $ is the spin-2 polarization tensor,
which is symmetric, $\epsilon^{\mu \nu}_\lambda = \epsilon^{\nu \mu}_\lambda$, 
traceless $g_{\mu \nu }\epsilon^{\mu \nu}_\lambda=0$, and transverse $q_\mu \epsilon^{\mu \nu}_\lambda =0 $. The extra
factor of ${1}/{3}$ in the scalar case is conventional, chosen such that $\langle S | \Theta| 0 \rangle = f_S q^2$. 
The {\em on-shell} couplings of the resonances to the $\pi\pi$ continuum are taken as
\begin{eqnarray}
  \langle S | \pi \pi \rangle = g_{S\pi\pi}, \;\;\;  \langle T | \pi \pi \rangle = g_{T\pi\pi} \epsilon^{\alpha \beta}_\lambda P^\alpha P^\beta  .
%   = g_{T\pi\pi} \epsilon^{\alpha \beta}_\lambda p'^\alpha p^\beta . \nonumber
\end{eqnarray}
Thus, we get
\begin{eqnarray}
&& \hspace{-7mm} \frac1{\pi }{\rm Im} \langle \pi \pi | \Theta^{\mu\nu} | 0\rangle = 
   \sum_S \frac{g_{S\pi\pi} f_S}3 \delta(m_S^2-q^2) m_S^2 Q^{\mu\nu}   \nonumber  \\ 
&& +  \sum_{T,\lambda}   \epsilon_\lambda^{\alpha \beta} P^\alpha P^\beta
  \epsilon^{\mu\nu}_\lambda g_{T\pi\pi} f_T  \delta(m_T^2-q^2),
\end{eqnarray}
which naturally complies with separate conservation for each term, yielding zero when contracted
with $q^{\mu}$. The sum over the tensor polarizations is given by~\cite{Scadron:1968zz,Novozhilov:1975yt},
\begin{eqnarray}
  \sum_{\lambda}   \epsilon_\lambda^{\alpha \beta} \epsilon^{\mu\nu}_\lambda 
= \frac12 \left( Q^{\mu \alpha} Q^{\nu \beta} + Q^{\nu \alpha} Q^{\mu \beta}
  \right) - \frac13 Q^{\mu \nu} Q^{\alpha \beta}.
   \end{eqnarray}
The on-shell condition $P \cdot q=0$ implies $P_\alpha Q^{\alpha,\beta}= P^\beta $, hence we obtain 
\begin{eqnarray}
\sum_{\lambda}   \epsilon_\lambda^{\alpha \beta} P_\alpha P_\beta \epsilon^{\mu\nu}_\lambda 
  = P^\mu P^\nu - \frac13 P^2 Q^{\mu \nu} 
   \end{eqnarray}
(cf. the tensor structure in Eq.~(\ref{eq:thetaS})). Therefore, in the narrow resonance, large-$N_c$ motivated approach we get 
\begin{eqnarray}
  \frac1{\pi}  {\rm Im} \,A (s) &=& \frac12     \sum_{T} g_{T\pi\pi} f_T  \delta( m_T^2-q^2) , \label{eq:del} \\
    \frac1{\pi}  {\rm Im} \,\Theta (s) 
  &=&  \sum_S g_{S\pi\pi} f_S m_S^2 \delta( m_S^2-q^2), \nonumber
  \end{eqnarray}
where, as expected, $A$ and $\Theta$ get contributions exclusively from the $2^{++}$ and $0^{++}$ states, respectively.  
As already suggested 
in~\cite{Donoghue:1991qv}, taking just one resonance per channel, i.e., using ${\rm Im}\,A(s) = \pi m_{f_2}^2 \delta\left ( s-m_{f_2}^2 \right )$ 
and ${\rm Im}\,\Theta(s) = \pi m_\sigma^4 \delta\left ( s-m_\sigma^2 \right )$ in the dispersion relations~(\ref{eq:A},\ref{eq:Th}), gives
\begin{eqnarray}
&& A(-Q^2)=\frac{m_{f_2}^2}{m_{f_2}^2+Q^2}, \label{eq:mf} \\
&& \Theta(-Q^2)=2m_\pi^2 - \frac{m_\sigma^2 Q^2}{m_\sigma^2+Q^2}, \label{eq:ms}
\end{eqnarray}
where the $2^{++}$ state is the $f_2(1270)$ meson, and the $0^{++}$
state is the $\sigma$ or $f_0(500)$ meson. We take here the usual
implementation of the short distance constraints, where only powers of
momenta are considered, neglecting running of $\alpha$~\cite{Pich:2002xy}. Equations (\ref{eq:mf},\ref{eq:ms}) correspond 
to taking $ g_{f_2\pi\pi} f_{f_2} = 2 m_{f_2}^2 $
and $ g_{\sigma \pi\pi} f_\sigma = m_\sigma^2$, which should be
understood as large-$N_c$  relations between {\em real} numbers within the 
minimal hadronic (narrow resonance) ansatz.\footnote{
%The truly resonance character as
%  poles and residues on the second Riemann sheet of these states make
  The parameters referring to the poles on the second Riemann sheet become complex at finite $N_c$.
  Recent compilations~\cite{Moussallam:2011zg,Hoferichter:2023mgy}
  yield (note an extra $\sqrt{2/3}$ factor there)  $|g_{\sigma
    \pi\pi} f_\sigma/ m_\sigma^2 | = 1.5(1) $ and $|g_{f_0 \pi\pi} f_{f_0}/
  m_{f_0}^2 | = 0.32(5) $, which shows the smallness of the $f_0(980)$
  contribution and differs from our % relations significantly but
 numbers as a (numerically significant) ${\cal O}(1/N_c)$ correction. However, what counts in
  practice is the dispersion integral with the spectral density along
  {\em real} $s$, including various ``background'' effects such as $\chi$PT or pQCD, and not just the residues at the complex poles. }
  
Now we are ready to use the lattice QCD data~\cite{Hackett:2023nkr}, available for $A$ and $D$,  to fit with the $\chi^2$ method 
the masses in Eqs.~(\ref{eq:mf},\ref{eq:ms}). We use 
relation (\ref{eq:Drel}) to express $D$ via formulas~(\ref{eq:mf},\ref{eq:ms}).
The result is
$m_{f_2}=1.24(3)~{\rm GeV}$ and $m_\sigma=0.65(3)~{\rm GeV}$, with
$\chi^2/{\rm DOF}=0.8$ for 49 data points. 
Note that these values correspond to the lattice pion mass $m_\pi=170$~MeV. The obtained $m_{f_2}$
is lower from the Particle Data Group central value of the BW mass, $m_{f_2}=1.275$~MeV (which is of course at the physical pion mass).
However, the effect is statistically not significant, as the the value of $\chi^2$ per point for the data of $A$ only is 0.54 for 
$m_{f_2}=1.24$~GeV and 0.7 for $m_{f_2}=1.275$~GeV. Moreover, adding more states in Eq.~(\ref{eq:del}) may change the fitted value of the single monopole mass.

Our results  are compared to the lattice data~\cite{Hackett:2023nkr} in Fig.~\ref{fig:fit}a.
A successful reproduction of $A(t)$ echoes the fit to the early
data of~\cite{Brommel:2007zz,QCDSF:2007ifr} in a similar model~\cite{Masjuan:2012sk}.
The agreement for $D(t)$ is equally satisfactory, with $D(0)=-1+4m_\pi^2/3m_{f_2}^2\simeq -0.97$. 
The gravitational mean squared radii, defined as $\langle r^2 \rangle_F = 6 F'(0)/F(0)$, are 
\begin{eqnarray}
&& \langle r^2 \rangle_A =\frac{6}{m_{f_2}^2} \simeq (0.39~{\rm fm})^2, \label{eq:radnar} \\
&& \langle r^2 \rangle_D = \frac{1}{D(0)} \left [ -\frac{2}{m_{f_2}^2} 
 - \frac{4}{m_\sigma^2} +\frac{8m_\pi^2}{m_{f_2}^4} \right ]\simeq (0.66~{\rm fm})^2, \nonumber
\end{eqnarray}
with the numbers corresponding to $m_\pi=170$~MeV. Substitution of the physical pion and $f_2$ masses to the above formulas yields
$D(0)=0.98$ and $ \langle r^2 \rangle_A =0.38$, a 1\% and 3\% effect, respectively. The radius of $D$, involving the $\sigma$, is discussed below.

In Fig.~\ref{fig:fit}b we present $\Theta$ from Eq.~(\ref{eq:ms}) with $m_\sigma=0.65(3)~{\rm GeV}$ as obtained above.
%To remove the trivial effect of  the pion mass at the value at the origin, we compare the subtracted quantity $\Theta(t)-\Theta(0)=\Theta(t)-2m_\pi^2$.
We can see that the
present narrow resonance model (the band) lies well within the data points obtained from~\cite{Hackett:2023nkr}  via relation~(\ref{eq:th0})
(we have added the errors in quadrature, which may not be accurate due to possible correlations). 

\begin{figure}[tb]
\begin{center}
\includegraphics[angle=0,width=0.38 \textwidth]{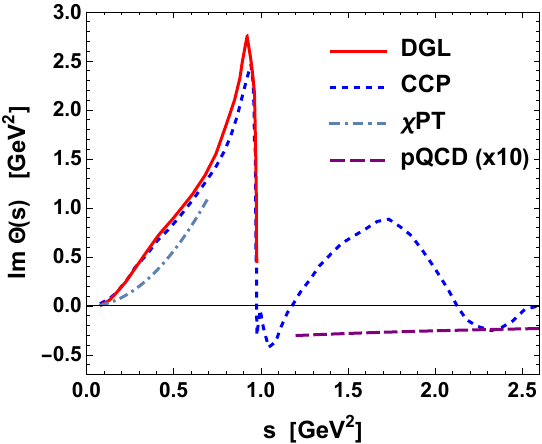} 
\end{center}
\vspace{-5mm}
\caption{Digitized ${\rm Im}\, \Theta(s)$ from Fig.~3, case $A_1$, of~\cite{Donoghue:1990xh} (DGL) and \cite{Celis:2013xja} (CCP). 
Note that the spectral density incorporates the $\pi\pi$ rescattering and $K \bar K$ effects, thus it has the physics of the 
$\sigma$ meson (a mild bump around $s\sim 0.4~{\rm GeV}$) and $f_0(980)$ (a minimum near its mass). For reference, we also plot 
the NLO $\chi$PT~\cite{Donoghue:1991qv} and the LO pQCD~\cite{Tong:2022zax} of Eq.~(\ref{eq:im}) (multiplied by 10).  
\label{fig:spect}} 
\end{figure}

{\bf Width effects.}  With the present lattice accuracy, any further
improvements of the model, such as adding higher states or nonzero
widths with new {fitted} parameters, is difficult to verify
numerically due to appearance of overfitting. The masses in monopole
fits to form factors are typically lower from the BW values, cf. the
case of the pion charge form factor~\cite{Masjuan:2012sk}.
Nevertheless, given the fact that $f_0(500)$ is very broad, it is important to attempt to analyze its width effects.  
In a more sophisticated treatment, ${\rm Im}\,\Theta(s)$ follows from a solution of  the coupled $\pi \pi$
and $K \bar{K}$ channel Omn\`es-Muskhelishvili equations~\cite{Pham:1976yi} in the scalar-isoscalar
channel~\cite{Truong:1989my,Donoghue:1990xh, Moussallam:2011zg, Celis:2013xja},
using the $\pi\pi$ and $K \bar K$ scattering data as input. Some additional information, in particular from $\chi$PT, is needed
to fix free constants appearing in the approach.
% previous analyses the physical scalar-isoscalar phase shifts are used as input to solve, with constants fixed by $\chi$PT.
Out of numerous calculations, we take two representative cases:
$A_1$ from Fig.~3
of~\cite{Donoghue:1990xh} (DGL), corresponding to a solution with the
CERN-Cracow-Munich~\cite{becker1979model} phase shifts as
input and from Fig.~4 from~\cite{Celis:2013xja} (CCP) implementing the Roy
equations solutions by the Bern~\cite{Ananthanarayan:2000ht} and
Madrid-Cracow~\cite{Garcia-Martin:2011iqs} groups.
The digitized ${\rm Im}\,\Theta(s)$ both cases are shown in Fig.~\ref{fig:spect}.  The
upper limits are $s_f \simeq 1 {\rm~GeV}^2$ for DGL and $s_f \simeq 3 {\rm~GeV}^2$ for CCP. In the vicinity of
$f_0(980)$  ${\rm Im}\, \Theta(s)$ sharply drops to a low value.

Then, we use the dispersion relation~(\ref{eq:Th}), integrating up to
$s_f$ to obtain $\Theta(-Q^2)$ in the space-like region. 
The results (obtained at $m_\pi=140$~MeV) are shown in Fig.~\ref{fig:fit}. We note that DGL and CCP are very close to each other and somewhat harder than the
narrow resonance fit (at $m_\pi=170$~MeV).
The model slope at the origin is 
\begin{eqnarray}
\left . d \Theta(t)/dt \right |_{t=0} = \frac{1}{\pi}  \int_{4 m_\pi^2}^{s_f} ds \frac{{\rm Im}\, \Theta (s)}{s^2}, \label{eq:Thsl}
\end{eqnarray}
which is $\simeq 1$ for both DGL and CCP, 
corresponding to the chiral limit behavior indicated in Fig.~\ref{fig:fit}b with the dashed
straight line. 
%The extended spectral analysis
%of \cite{Celis:2013xja} up to $s=3 {\rm GeV}^2$ (see
%Fig.~\ref{fig:spect}), preserves these features suggesting a large
%cancellation in the spectral integral above the $f_0(980)$ mass.
%
%We note that the model follows the data within the uncertainties.
At the origin
\begin{eqnarray}
D(0) = -1 +\frac{4m_\pi^2}{3m_{f_2}^2}-\frac{2}{3} \left [  \left . d \Theta(t)/dt \right |_{t=0} -1  \right ], 
\end{eqnarray}
which yields 0.98 for DGL and 0.96 for CCP.
The corresponding radius is 
\begin{eqnarray}
 \langle r^2 \rangle_D = \frac{1}{D(0)} \left [ -\frac{2}{m_{f_2}^2} 
 - \frac{4}{\pi} \int_{4 m_\pi^2}^{s_f} ds \frac{{\rm Im} \,\Theta (s)}{s^3} +\frac{8m_\pi^2}{m_{f_2}^4} \right ], \label{eq:radii} 
\end{eqnarray}
giving 0.71 for DGL and 0.70 for CCP.
The obtained numbers follow the hierarchy pattern $\langle r^2 \rangle_A  < \langle r^2 \rangle_{EM} <\langle r^2 \rangle_D$ (where 
$\langle r^2 \rangle_{EM}=(0.659(4)~{\rm fm})^2$ is the electromagnetic ms radius of the pion) are consistent with 
the analysis of~\cite{Kumano:2017lhr} for the quark radii (our value for $\langle r^2 \rangle_D$ is 20\% smaller).
%
%This feature is in fact consistent
%with the expected insensitivity of the space-like physics to the
%time-like details.
%
We also remark that the lowest $Q^2=-t$ data point for $\Theta(t)$ is above zero at a $2.5~\sigma$ level, in 
accordance with the positivity of $\Theta(0)=2m_\pi^2$. 

At larger values of $Q^2$ the errors are too large to draw strong conclusions, yet in the covered range the data seem to flatten. 
At asymptotic $Q^2$, Eq.~(\ref{eq:Th}) with the literally taken
spectral functions of Fig.~\ref{fig:spect} and the upper limit $s_f$ gives
\begin{eqnarray}
\Theta(-Q^2) \sim 2m_\pi^2-\int_{4 m_\pi^2}^{s_f} \!\!\! ds \, \frac{{\rm Im}\, \Theta (s)}{s}, \label{eq:msr}
\end{eqnarray}
which yields negative $\Theta(-Q^2) \simeq 2m_\pi^2-0.49~{\rm GeV}^2$ for DGL and $2m_\pi^2-0.52~{\rm GeV}^2$ for CCP.
%In the chiral limit, sum rule~(\ref{eq:msr}) with a positive spectral function means that the large-$Q^2$ value of $\Theta$ is negative. 
Given the smallness of $\Theta(0)=2 m_\pi^2$ and the slope $\simeq 1$, one expects a zero at 
$t=-Q_0^2$, i.e., $\Theta(-Q_0^2)=0$, where at LO in the chiral expansion $Q_0^2= 2 m_\pi^2$.  This behavior is indeed seen 
in the data, where the change of sign  occurs near $0.07~{\rm GeV}^2$, slightly above the lattice value of $2m_\pi^2$. 

{\bf Pion mass effects.} The dependence on  $m_\pi$  has been 
studied in the literature for the corresponding poles in the second Riemann sheet,  
  which due to unitarity is shared by the form factors and the
  corresponding scattering amplitudes (see, e.g., ~\cite{Nieves:1999bx}). For instance, within 
  the inverse amplitude method~\cite{Hanhart:2008mx} (based on $\chi$PT),
  changing $m_\pi$ from the physical point up to 170~MeV, 
%  in $\sqrt{s_\sigma}= M_\sigma- i \Gamma_\sigma/2 $, 
  increases the $\sigma$ pole mass by 2-3\% and decreases its width by
  5-8$\%$.
%  This general trend is reproduced also by a quite different (not based on $\chi$PT) method~\cite{Rupp:2024tyh}.

\begin{figure}[tb]
\begin{center}
\includegraphics[angle=0,width=0.4 \textwidth]{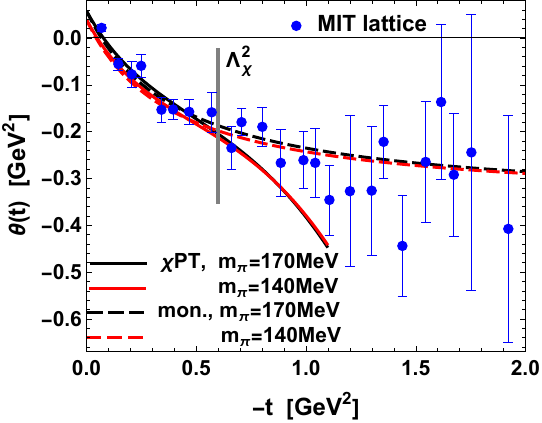} 
\end{center}
\vspace{-5mm}
\caption{The trace anomaly form factor of the pion, plotted as a function of the space-like momentum
  transfer $-t=Q^2$: Lattice QCD data from~\cite{Hackett:2023nkr}
  (points), $\chi$PT result~\cite{Donoghue:1990xh} from Eq.~(\ref{eq:ThetaChPT}), 
  fitted at the lattice pion mass
  as well as the physical point (solid lines). For both cases the corresponding monopole fits with
  $m_\sigma=0.64(3)$~GeV and $m_\sigma=0.63(6)$~GeV are also displayed (dashed lines). 
    \label{fig:ChPTfit} }
\end{figure}

  Here, however, we are after the change with $m_\pi$ of the {\it monopole} mass, which only
  agrees with both the BW and the pole masses in the large $N_c$
  limit. In $\chi$PT (at NLO) the  GFFs~\cite{Donoghue:1990xh}  can suitably be
  written as
\begin{eqnarray}
\Theta (t) &=& 2 m_\pi^2 + t - \frac{\bar c_1 m_\pi^2 t /2
  - \bar c_2 t^2}{(4 \pi f_\pi)^2} \nonumber \\
&+& \frac{t^3}{\pi} \int_{4 m_\pi^2}^\infty \frac{ds}{s^3}\frac{{\rm Im} \Theta(s)}{s-t} + {\cal O} (f_\pi^{-4}) \label{eq:ThetaChPT} \\ 
  A(t) &=& 1 - \frac{2 L_{12}}{f_\pi^2} t  + {\cal O} (f_\pi^{-4})
\label{eq:AChPT}
\end{eqnarray}
where $\bar c_i = c_i^r (\mu) + \ln (\mu^2/m_\pi^2)$
encode the $\chi$PT low energy coefficients as
% $L_{11,12,13}$ LECs as 
  $  c_1^r=1-128 \pi^2 (6 L_{11}+L_{12}-6 L_{13})$ and 
  $  c_2^r=11/10-64 \pi^2 (3 L_{11}+L_{12})$ (the $K \bar K$ and $\eta \eta$ threshold effects are neglected). 
The spectral function is~\cite{Donoghue:1990xh} 
\begin{eqnarray}
  && \hspace{-7mm} \frac1{\pi} {\rm Im}\, \Theta (s) =\sqrt{1-\frac{4 m_\pi^2}{s}} \frac{\left(2 m_\pi^2+s\right) \left(2 s-m_\pi^2\right)}{32 \pi ^2 f_\pi^2}
\label{eq:ImThetaChPT}
\end{eqnarray}
% ** which is manifestly positive 
(see Fig.~\ref{fig:spect}). 
Taking $\mu=m_\rho $ (independent of $m_\pi$) and fitting the lattice data in the $\chi$PT fiducial region $Q^2 \le \mu^2$
at $m_\pi=0.170$~GeV (and hence $f_\pi=0.095$~GeV) yields 
\begin{eqnarray}
 c_1^r = -10(7)\, , \quad c_2^r = 0.46(28), 
\end{eqnarray}  
with a satisfactory $\chi^2 /{\rm DOF}= 1.5 < 1 +\sqrt{2/{\rm  DOF}}=1.57$ for DOF=8-2=6 (see Fig.~\ref{fig:ChPTfit}). The strong
correlation $\rho(c_1^r,c_2^r) = -0.95$ indicates effectively one parameter. Indeed, 
fitting the $\chi$PT curve with the monopole form (\ref{eq:ms}) for $Q^2 \le m_\rho^2$ yields $m_\sigma = 0.64(3)$~GeV, in agreement with the value $0.65(3)$~GeV
from the previous fit to all lattice data  (cf. Fig.~\ref{fig:fit}).

Having fixed the constants $c_i^r (\mu)$ we may use Eq.~(\ref{eq:ThetaChPT}) to promptly
extrapolate to the physical point ($m_\pi=140$~MeV, $f_\pi=0.093$~GeV).
The result is shown in Fig.~\ref{fig:ChPTfit}, where we note the proximity between the lattice and physical pion masses. 
The corresponding monopole fit gives $m_\sigma=0.63(6)$~GeV, thus the pion mass effects are at about
$2\%$ level (see Fig.~\ref{fig:ChPTfit}). The DGL and CPP $\pi \pi$
and $K \bar{K}$ coupled channel unitarized calculations cited above can be approximately described by a
monopole with $m_\sigma=0.68$~GeV, consistent within uncertainties with the monopole fit to $\chi$PT at the physical point.

Likewise, Eq.~(\ref{eq:AChPT}) suggests that $m_{f_2}$ is approximately proportional to $f_\pi $, hence
it is expected that the value of $m_{f_2}$ should mildly grow with
$m_\pi$. On the lattice one has $m_{f_2}=1470$~MeV for
$m_\pi=391$~MeV~\cite{Briceno:2017qmb}.

{\bf Anatomy of mass sum rule.} The dispersion relation~(\ref{eq:Th}) at asymptotic $Q^2$, together with the pQCD limit, yields
\begin{eqnarray}
2 m_\pi^2 - \frac1{\pi} \int_{4 m_\pi^2}^\infty ds \frac{{\rm Im}\, \Theta (s)}{s}=0. \label{eq:th2}
\end{eqnarray}
We will consider the low- and high $s$ contributions to this sum rule to draw conclusions on the spectral density. Firstly, 
from NLO $\chi$PT one has~\cite{Donoghue:1990xh}
Eq.~(\ref{eq:ImThetaChPT}).  Taking
%, for the sake of an estimate,
$\Lambda_\chi  
% = 4 m_\pi 
\sim 0.6$~GeV, 
% corresponding to the $4\pi$ production threshold, 
one obtains the contribution 
% to the mass sum rule~(\ref{eq:msr})
\begin{eqnarray}
  -\frac1{\pi} \int_{4 m_\pi^2}^{\Lambda_\chi^2} ds \frac{{\rm Im} \,\Theta (s)}s %= - 0.66 \frac{m^4}{f^2}
  = - 0.03~{\rm GeV}^2, %+ {\cal O} (p^4) ,
\end{eqnarray}
which is comparable in size to  $\Theta(0)=2m_\pi^2$, 
and tiny compared to the values reached in Fig.~\ref{fig:fit}b at higher $Q^2$.
In view of the earlier discussion, where the integration with the spectral densities of Fig.~\ref{fig:spect} up to 
$s_f \sim1-3~{\rm GeV}^2$ led to large negative values of $\Theta(-Q^2)$ at large $Q^2$, Eq.~(\ref{eq:th2}) means that 
${\rm Im} \, \Theta (s)$ 
% cannot be positive definite and needs to acquire
must acquire negative sign contributions at larger values of $s$ than the range of Fig.~\ref{fig:spect}
%, beyond the range shown in Fig.~\ref{fig:spect}. 
(we remark that positivity of a three-point spectral function is not formally 
protected). This situation resembles the case of the pion charge form factor, where a similar argument applies~\cite{Donoghue:1996bt}.
Actually, an analysis to higher $\sqrt{s} \le 3$~GeV shows that 
the charge spectral density changes signs at about $\sqrt{s} \sim 1.25,1.6,1.9, \dots~{\rm GeV}$~\cite{RuizArriola:2024gwb}.

At the other end, the dispersive integral can be evaluated upwards from a high mass
scale $\Lambda$ where pQCD sets in, by analytically
continuing the LO pQCD result from negative space-like region $t=-Q^2$
to the complex plane $-t= e^{-i \theta} |t|$.  The positive
time-like region corresponds to $\theta \to -\pi $, such that one has
$q^2= s+ i \epsilon $
and $\ln (Q^2/\Lambda_{\rm QCD}^2) \to \ln (s e^{-i \pi}/\Lambda_{\rm QCD}^2) = \ln (s/\Lambda_{\rm QCD}^2) - i \pi$. 
% On the upper lip of the cut, 
% ** at $ s \gg4 m_\pi^2 $,
With the notation $L= \log (s/\Lambda_{\rm QCD}^2) $, one gets
\begin{eqnarray}
\alpha(s) \equiv \alpha(s+ i \epsilon)= \left(\frac{4 \pi}{\beta_0} \right) \frac1{L  - i \pi}, 
\end{eqnarray}
yielding a positive imaginary part $ {\rm Im} \, \alpha (s+ i \epsilon)^2
= (4 \pi /\beta_0)^2 2 \pi L /(L^2+\pi^2)^2 $, hence 
\begin{eqnarray}
\frac1{\pi}  {\rm Im}\, \Theta (s) =  - \left(\frac{4 \pi}{\beta_0}\right)^2 \frac{8  \beta_0 L f_\pi^2}{(L^2+\pi^2)^2} + {\cal O} (\alpha^3). \label{eq:im}
\end{eqnarray}
This implies that $ {\rm Im}\, \Theta (s) < 0 $ at large energies, which is the desired result (see Fig.~\ref{fig:spect}). 
After computing the integral, the contribution to sum rule~(\ref{eq:th2}) becomes 
\begin{eqnarray}
 - \frac1{\pi} \int_{\Lambda^2}^\infty ds \frac{{\rm Im} \,\Theta (s)}{s} =  4 \beta_0 |\alpha (-\Lambda^2)|^2 f_\pi^2 + {\cal O} (\alpha^3),
\end{eqnarray}
which at 
% even the lowest scale 
$\Lambda^2 \!\sim \! 20 ~\Lambda_{\rm
  QCD}^2 \!\sim \! 1~{\rm GeV}^2 $ is about $0.03~{\rm GeV}^2$, small
compared to the values reached at high $Q^2$ in Fig.~\ref{fig:fit}.
%\footnote{The power-suppressed quark contribution to the sum rule appears much smaller, 
%$ - \frac1{\pi} \int_{\Lambda^2}^\infty  ds \frac{1}{s} {\rm Im \Theta_q (s)}=  c \times  0.3 \times 10^{-4}{\rm GeV}^2$.}
This means that the spectral function must acquire larger negative values at intermediate values of $s$, 
%at lower values of $s$ than $\Lambda^2$, and/or 
for instance pick them up from higher order pQCD or other non-perturbative effects. 

{\bf More sum rules.} The asymptotic values of
$A(-Q^2)$ and $D(-Q^2)$ vanish as $1/Q^2$ divided by the log corrections~\cite{Tong:2021ctu,Tong:2022zax,Krutov:2023ztx}.
Therefore one has:  
%the following sum rules immediately follow:
\begin{eqnarray}
  0 &=& 1-\frac1{\pi} \int_{4 m_\pi^2}^\infty ds \frac{1}{s}{\rm Im} \,A(s), \nonumber  \\
  0 &=& D(0)-\frac1{\pi} \int_{4 m_\pi^2}^\infty ds \frac{1}{s} {\rm Im} \,D(s),   .
\end{eqnarray}
In fact, since $Q^2 A(-Q^2)$ and $Q^2 D(-Q^2)$ also tend to zero due to the extra
 $~\alpha\sim 1/\ln Q^2$ suppression one has 
%by  from $\alpha$, 
\begin{eqnarray}
  0 = \frac1{\pi} \int_{4 m_\pi^2}^\infty ds \,{\rm Im} \,A(s), \;\;\;\; 0 = \frac1{\pi} \int_{4 m_\pi^2}^\infty ds \,{\rm Im} \,D(s),
\end{eqnarray}
implying in particular that the spectral densities for $A(s)$, $D(s)$  cannot have a well-defined sign. For  ${\rm Im} \Theta(s)$, since 
for $1/2 < \Theta'(0)= 1 + {\cal O} (m_\pi^2/f_\pi^2)$, one has 
\begin{eqnarray}
2 m_\pi^2 (1\!-\!2 \Theta'(0))= \frac1{\pi} \int_{4 m_\pi^2}^\infty \!\!\! ds  (s-4m_\pi^2)  \frac{{\rm Im} \,\Theta(s)}{s^2} < 0 \, , 
\end{eqnarray}
so ${\rm Im} \Theta(s)$ must change sign, as 
illustrated explicitly by the alternating pattern of the CCP case in
Fig.~\ref{fig:spect}.

{\bf Conclusions.} The recent
% high precision 
MIT lattice QCD data for the gravitational form factors of the pion can be naturally understood and 
accurately described with {\em meson dominance}, working properly in the momentum transfer range covered by the 
lattice. The analysis requires the projection on good spin quantum numbers. The tensor $2^{++}$ component, corresponding to $A$, can be 
saturated with the $f_2(1270)$ meson, whereas the scalar $0^{++}$ component (the trace anomaly form factor $\Theta$), similarly, 
with the $\sigma$ meson. More accurate description is achieved by using the 
physical spectral function in the dispersion relation. 
The form factor $D$ is obtained as a combination of $A$ and $\Theta$.
We have discussed possible effects of $m_\pi$ between the physical point at the lattice value of 170~MeV, with the conclusion that they are small.
We have also exploited the pQCD constraints at high momentum transfers, with the peculiar feature that at 
infinite space-like momentum  $\Theta(-Q^2)$ goes to a constant times $1/\log^2 Q^2$ corrections. These constraints lead to 
sum rules for the spectral densities of the gravitational form factors, which imply that these densities cannot be of definite sign.
The numerical smallness of the (negative) LO pQCD contribution to the spectral density of $\Theta$, together with a positive contribution from $\chi$PT and
a large positive contribution from the 
resonance region up to $s \sim 1~{\rm GeV}^2$, mean that the higher mass resonances or higher orders in pQCD must bring in large negative
contributions to the spectral density of $\Theta$. This feature is needed to reconcile the available lattice data and the theoretical requirements.

\medskip

We are grateful to the authors of Ref.~\cite{Hackett:2023nkr} for providing us the data tables for Fig.~\ref{fig:fit}a, as well as 
to the authors of Ref.~\cite{Wang:2024lrm} for communicating their numerical results. We thank Pablo S\'anchez Puertas for discussions.   
Supported by NCN grant 2018/31/B/ST2/01022 (WB), by the Spanish MINECO and European FEDER funds grant
and Project No. PID2020–114767 GB-I00 funded by MCIN/AEI/10.13039/501100011\-033, and by the Junta de Andaluc{\'i}a grant FQM-225 (ERA).

\bibliography{newrefs,refs-barpi,Ref}

\begin{thebibliography}{56}
\expandafter\ifx\csname natexlab\endcsname\relax\def\natexlab#1{#1}\fi
\providecommand{\url}[1]{\texttt{#1}}
\providecommand{\href}[2]{#2}
\providecommand{\path}[1]{#1}
\providecommand{\DOIprefix}{doi:}
\providecommand{\ArXivprefix}{arXiv:}
\providecommand{\URLprefix}{URL: }
\providecommand{\Pubmedprefix}{pmid:}
\providecommand{\doi}[1]{\href{http://dx.doi.org/#1}{\path{#1}}}
\providecommand{\Pubmed}[1]{\href{pmid:#1}{\path{#1}}}
\providecommand{\bibinfo}[2]{#2}
\ifx\xfnm\relax \def\xfnm[#1]{\unskip,\space#1}\fi
%Type = Article
\bibitem[{Hackett et~al.(2023)Hackett, Oare, Pefkou, and
  Shanahan}]{Hackett:2023nkr}
\bibinfo{author}{D.~C. Hackett}, \bibinfo{author}{P.~R. Oare},
  \bibinfo{author}{D.~A. Pefkou}, \bibinfo{author}{P.~E. Shanahan},
\newblock \bibinfo{title}{{Gravitational form factors of the pion from lattice
  QCD}},
\newblock \bibinfo{journal}{Phys. Rev. D} \bibinfo{volume}{108}
  (\bibinfo{year}{2023}) \bibinfo{pages}{114504}.
  \DOIprefix\doi{10.1103/PhysRevD.108.114504}.
  \href{http://arxiv.org/abs/2307.11707}{{\tt arXiv:2307.11707}}.
%Type = Phdthesis
\bibitem[{Pefkou(2023)}]{Pefkou:2023okb}
\bibinfo{author}{D.~A. Pefkou}, \bibinfo{title}{{Gravitational form factors of
  hadrons from lattice QCD}}, Ph.D. thesis, MIT, \bibinfo{year}{2023}.
%Type = Phdthesis
\bibitem[{{Br\"{o}mmel, Dirk}(2007)}]{Brommel:2007zz}
\bibinfo{author}{{Br\"{o}mmel, Dirk}}, \bibinfo{title}{{Pion Structure from the
  Lattice}}, Ph.D. thesis, Regensburg U., \bibinfo{year}{2007}.
  \DOIprefix\doi{10.3204/DESY-THESIS-2007-023}.
%Type = Article
\bibitem[{Br\"ommel et~al.(2008)}]{QCDSF:2007ifr}
\bibinfo{author}{D.~Br\"ommel}, et~al. (\bibinfo{collaboration}{QCDSF, UKQCD}),
\newblock \bibinfo{title}{{The Spin structure of the pion}},
\newblock \bibinfo{journal}{Phys. Rev. Lett.} \bibinfo{volume}{101}
  (\bibinfo{year}{2008}) \bibinfo{pages}{122001}.
  \DOIprefix\doi{10.1103/PhysRevLett.101.122001}.
  \href{http://arxiv.org/abs/0708.2249}{{\tt arXiv:0708.2249}}.
%Type = Inproceedings
\bibitem[{Delmar et~al.(2024)Delmar, Alexandrou, Bacchio, Clo\"et,
  Constantinou, and Koutsou}]{Delmar:2024vxn}
\bibinfo{author}{J.~Delmar}, \bibinfo{author}{C.~Alexandrou},
  \bibinfo{author}{S.~Bacchio}, \bibinfo{author}{I.~Clo\"et},
  \bibinfo{author}{M.~Constantinou}, \bibinfo{author}{G.~Koutsou},
\newblock \bibinfo{title}{{Generalized form factors of the pion and kaon using
  twisted mass fermions}},
\newblock in: \bibinfo{booktitle}{{40th International Symposium on Lattice
  Field Theory}}, \bibinfo{year}{2024}.
  \href{http://arxiv.org/abs/2401.04080}{{\tt arXiv:2401.04080}}.
%Type = Article
\bibitem[{Shanahan and Detmold(2019)}]{Shanahan:2018pib}
\bibinfo{author}{P.~E. Shanahan}, \bibinfo{author}{W.~Detmold},
\newblock \bibinfo{title}{{Gluon gravitational form factors of the nucleon and
  the pion from lattice QCD}},
\newblock \bibinfo{journal}{Phys. Rev. D} \bibinfo{volume}{99}
  (\bibinfo{year}{2019}) \bibinfo{pages}{014511}.
  \DOIprefix\doi{10.1103/PhysRevD.99.014511}.
  \href{http://arxiv.org/abs/1810.04626}{{\tt arXiv:1810.04626}}.
%Type = Article
\bibitem[{Wang et~al.(2024)Wang, He, Wang, Draper, Liang, Liu, and
  Yang}]{Wang:2024lrm}
\bibinfo{author}{B.~Wang}, \bibinfo{author}{F.~He}, \bibinfo{author}{G.~Wang},
  \bibinfo{author}{T.~Draper}, \bibinfo{author}{J.~Liang},
  \bibinfo{author}{K.-F. Liu}, \bibinfo{author}{Y.-B. Yang}
  (\bibinfo{collaboration}{\ensuremath{\chi}QCD}),
\newblock \bibinfo{title}{{Trace anomaly form factors from lattice QCD}},
\newblock \bibinfo{journal}{Phys. Rev. D} \bibinfo{volume}{109}
  (\bibinfo{year}{2024}) \bibinfo{pages}{094504}.
  \DOIprefix\doi{10.1103/PhysRevD.109.094504}.
  \href{http://arxiv.org/abs/2401.05496}{{\tt arXiv:2401.05496}}.
%Type = Article
\bibitem[{Pagels(1966)}]{PhysRev.144.1250}
\bibinfo{author}{H.~Pagels},
\newblock \bibinfo{title}{Energy-momentum structure form factors of particles},
\newblock \bibinfo{journal}{Phys. Rev.} \bibinfo{volume}{144}
  (\bibinfo{year}{1966}) \bibinfo{pages}{1250--1260}.
  \DOIprefix\doi{10.1103/PhysRev.144.1250}.
%Type = Article
\bibitem[{Polyakov and Schweitzer(2018)}]{Polyakov:2018zvc}
\bibinfo{author}{M.~V. Polyakov}, \bibinfo{author}{P.~Schweitzer},
\newblock \bibinfo{title}{{Forces inside hadrons: pressure, surface tension,
  mechanical radius, and all that}},
\newblock \bibinfo{journal}{Int. J. Mod. Phys. A} \bibinfo{volume}{33}
  (\bibinfo{year}{2018}) \bibinfo{pages}{1830025}.
  \DOIprefix\doi{10.1142/S0217751X18300259}.
  \href{http://arxiv.org/abs/1805.06596}{{\tt arXiv:1805.06596}}.
%Type = Article
\bibitem[{Broniowski et~al.(2008)Broniowski, Ruiz~Arriola, and
  Golec-Biernat}]{Broniowski:2007si}
\bibinfo{author}{W.~Broniowski}, \bibinfo{author}{E.~Ruiz~Arriola},
  \bibinfo{author}{K.~Golec-Biernat},
\newblock \bibinfo{title}{{Generalized parton distributions of the pion in
  chiral quark models and their QCD evolution}},
\newblock \bibinfo{journal}{Phys. Rev. D} \bibinfo{volume}{77}
  (\bibinfo{year}{2008}) \bibinfo{pages}{034023}.
  \DOIprefix\doi{10.1103/PhysRevD.77.034023}.
  \href{http://arxiv.org/abs/0712.1012}{{\tt arXiv:0712.1012}}.
%Type = Article
\bibitem[{Broniowski and Ruiz~Arriola(2008)}]{Broniowski:2008hx}
\bibinfo{author}{W.~Broniowski}, \bibinfo{author}{E.~Ruiz~Arriola},
\newblock \bibinfo{title}{{Gravitational and higher-order form factors of the
  pion in chiral quark models}},
\newblock \bibinfo{journal}{Phys. Rev. D} \bibinfo{volume}{78}
  (\bibinfo{year}{2008}) \bibinfo{pages}{094011}.
  \DOIprefix\doi{10.1103/PhysRevD.78.094011}.
  \href{http://arxiv.org/abs/0809.1744}{{\tt arXiv:0809.1744}}.
%Type = Article
\bibitem[{Frederico et~al.(2009)Frederico, Pace, Pasquini, and
  Salme}]{Frederico:2009fk}
\bibinfo{author}{T.~Frederico}, \bibinfo{author}{E.~Pace},
  \bibinfo{author}{B.~Pasquini}, \bibinfo{author}{G.~Salme},
\newblock \bibinfo{title}{{Pion Generalized Parton Distributions with covariant
  and Light-front constituent quark models}},
\newblock \bibinfo{journal}{Phys. Rev. D} \bibinfo{volume}{80}
  (\bibinfo{year}{2009}) \bibinfo{pages}{054021}.
  \DOIprefix\doi{10.1103/PhysRevD.80.054021}.
  \href{http://arxiv.org/abs/0907.5566}{{\tt arXiv:0907.5566}}.
%Type = Article
\bibitem[{Masjuan et~al.(2013)Masjuan, Ruiz~Arriola, and
  Broniowski}]{Masjuan:2012sk}
\bibinfo{author}{P.~Masjuan}, \bibinfo{author}{E.~Ruiz~Arriola},
  \bibinfo{author}{W.~Broniowski},
\newblock \bibinfo{title}{{Meson dominance of hadron form factors and
  large-$N_c$ phenomenology}},
\newblock \bibinfo{journal}{Phys. Rev. D} \bibinfo{volume}{87}
  (\bibinfo{year}{2013}) \bibinfo{pages}{014005}.
  \DOIprefix\doi{10.1103/PhysRevD.87.014005}.
  \href{http://arxiv.org/abs/1210.0760}{{\tt arXiv:1210.0760}}.
%Type = Article
\bibitem[{Fanelli et~al.(2016)Fanelli, Pace, Romanelli, Salme, and
  Salmistraro}]{Fanelli:2016aqc}
\bibinfo{author}{C.~Fanelli}, \bibinfo{author}{E.~Pace},
  \bibinfo{author}{G.~Romanelli}, \bibinfo{author}{G.~Salme},
  \bibinfo{author}{M.~Salmistraro},
\newblock \bibinfo{title}{{Pion Generalized Parton Distributions within a fully
  covariant constituent quark model}},
\newblock \bibinfo{journal}{Eur. Phys. J. C} \bibinfo{volume}{76}
  (\bibinfo{year}{2016}) \bibinfo{pages}{253}.
  \DOIprefix\doi{10.1140/epjc/s10052-016-4101-1}.
  \href{http://arxiv.org/abs/1603.04598}{{\tt arXiv:1603.04598}}.
%Type = Article
\bibitem[{Freese and Clo\"et(2019)}]{Freese:2019bhb}
\bibinfo{author}{A.~Freese}, \bibinfo{author}{I.~C. Clo\"et},
\newblock \bibinfo{title}{{Gravitational form factors of light mesons}},
\newblock \bibinfo{journal}{Phys. Rev. C} \bibinfo{volume}{100}
  (\bibinfo{year}{2019}) \bibinfo{pages}{015201}.
  \DOIprefix\doi{10.1103/PhysRevC.100.015201}.
  \href{http://arxiv.org/abs/1903.09222}{{\tt arXiv:1903.09222}},
  \bibinfo{note}{[Erratum: Phys.Rev.C 105, 059901 (2022)]}.
%Type = Article
\bibitem[{Krutov and Troitsky(2021)}]{Krutov:2020ewr}
\bibinfo{author}{A.~F. Krutov}, \bibinfo{author}{V.~E. Troitsky},
\newblock \bibinfo{title}{{Pion gravitational form factors in a relativistic
  theory of composite particles}},
\newblock \bibinfo{journal}{Phys. Rev. D} \bibinfo{volume}{103}
  (\bibinfo{year}{2021}) \bibinfo{pages}{014029}.
  \DOIprefix\doi{10.1103/PhysRevD.103.014029}.
  \href{http://arxiv.org/abs/2010.11640}{{\tt arXiv:2010.11640}}.
%Type = Article
\bibitem[{Xing et~al.(2023)Xing, Ding, and Chang}]{Xing:2022mvk}
\bibinfo{author}{Z.~Xing}, \bibinfo{author}{M.~Ding},
  \bibinfo{author}{L.~Chang},
\newblock \bibinfo{title}{{Glimpse into the pion gravitational form factor}},
\newblock \bibinfo{journal}{Phys. Rev. D} \bibinfo{volume}{107}
  (\bibinfo{year}{2023}) \bibinfo{pages}{L031502}.
  \DOIprefix\doi{10.1103/PhysRevD.107.L031502}.
  \href{http://arxiv.org/abs/2211.06635}{{\tt arXiv:2211.06635}}.
%Type = Article
\bibitem[{Xu et~al.(2024)Xu, Ding, Raya, Roberts, Rodr\'\i{}guez-Quintero, and
  Schmidt}]{Xu:2023izo}
\bibinfo{author}{Y.-Z. Xu}, \bibinfo{author}{M.~Ding},
  \bibinfo{author}{K.~Raya}, \bibinfo{author}{C.~D. Roberts},
  \bibinfo{author}{J.~Rodr\'\i{}guez-Quintero}, \bibinfo{author}{S.~M.
  Schmidt},
\newblock \bibinfo{title}{{Pion and kaon electromagnetic and gravitational form
  factors}},
\newblock \bibinfo{journal}{Eur. Phys. J. C} \bibinfo{volume}{84}
  (\bibinfo{year}{2024}) \bibinfo{pages}{191}.
  \DOIprefix\doi{10.1140/epjc/s10052-024-12518-x}.
  \href{http://arxiv.org/abs/2311.14832}{{\tt arXiv:2311.14832}}.
%Type = Article
\bibitem[{Li and Vary(2024)}]{Li:2023izn}
\bibinfo{author}{Y.~Li}, \bibinfo{author}{J.~P. Vary},
\newblock \bibinfo{title}{{Stress inside the pion in holographic light-front
  QCD}},
\newblock \bibinfo{journal}{Phys. Rev. D} \bibinfo{volume}{109}
  (\bibinfo{year}{2024}) \bibinfo{pages}{L051501}.
  \DOIprefix\doi{10.1103/PhysRevD.109.L051501}.
  \href{http://arxiv.org/abs/2312.02543}{{\tt arXiv:2312.02543}}.
%Type = Article
\bibitem[{Liu et~al.(2024{\natexlab{a}})Liu, Shuryak, Weiss, and
  Zahed}]{Liu:2024jno}
\bibinfo{author}{W.-Y. Liu}, \bibinfo{author}{E.~Shuryak},
  \bibinfo{author}{C.~Weiss}, \bibinfo{author}{I.~Zahed},
\newblock \bibinfo{title}{{Pion gravitational form factors in the QCD instanton
  vacuum. I}},
\newblock \bibinfo{journal}{Phys. Rev. D} \bibinfo{volume}{110}
  (\bibinfo{year}{2024}{\natexlab{a}}) \bibinfo{pages}{054021}.
  \DOIprefix\doi{10.1103/PhysRevD.110.054021}.
  \href{http://arxiv.org/abs/2405.14026}{{\tt arXiv:2405.14026}}.
%Type = Article
\bibitem[{Liu et~al.(2024{\natexlab{b}})Liu, Shuryak, and Zahed}]{Liu:2024vkj}
\bibinfo{author}{W.-Y. Liu}, \bibinfo{author}{E.~Shuryak},
  \bibinfo{author}{I.~Zahed},
\newblock \bibinfo{title}{{Pion gravitational form factors in the QCD instanton
  vacuum. II}},
\newblock \bibinfo{journal}{Phys. Rev. D} \bibinfo{volume}{110}
  (\bibinfo{year}{2024}{\natexlab{b}}) \bibinfo{pages}{054022}.
  \DOIprefix\doi{10.1103/PhysRevD.110.054022}.
  \href{http://arxiv.org/abs/2405.16269}{{\tt arXiv:2405.16269}}.
%Type = Article
\bibitem[{Kumano et~al.(2018)Kumano, Song, and Teryaev}]{Kumano:2017lhr}
\bibinfo{author}{S.~Kumano}, \bibinfo{author}{Q.-T. Song},
  \bibinfo{author}{O.~V. Teryaev},
\newblock \bibinfo{title}{{Hadron tomography by generalized distribution
  amplitudes in pion-pair production process $\gamma^* \gamma \rightarrow \pi^0
  \pi^0 $ and gravitational form factors for pion}},
\newblock \bibinfo{journal}{Phys. Rev. D} \bibinfo{volume}{97}
  (\bibinfo{year}{2018}) \bibinfo{pages}{014020}.
  \DOIprefix\doi{10.1103/PhysRevD.97.014020}.
  \href{http://arxiv.org/abs/1711.08088}{{\tt arXiv:1711.08088}}.
%Type = Article
\bibitem[{Ji(1995)}]{Ji:1994av}
\bibinfo{author}{X.-D. Ji},
\newblock \bibinfo{title}{{A QCD analysis of the mass structure of the
  nucleon}},
\newblock \bibinfo{journal}{Phys. Rev. Lett.} \bibinfo{volume}{74}
  (\bibinfo{year}{1995}) \bibinfo{pages}{1071--1074}.
  \DOIprefix\doi{10.1103/PhysRevLett.74.1071}.
  \href{http://arxiv.org/abs/hep-ph/9410274}{{\tt arXiv:hep-ph/9410274}}.
%Type = Article
\bibitem[{Lorc\'e(2018)}]{Lorce:2017xzd}
\bibinfo{author}{C.~Lorc\'e},
\newblock \bibinfo{title}{{On the hadron mass decomposition}},
\newblock \bibinfo{journal}{Eur. Phys. J. C} \bibinfo{volume}{78}
  (\bibinfo{year}{2018}) \bibinfo{pages}{120}.
  \DOIprefix\doi{10.1140/epjc/s10052-018-5561-2}.
  \href{http://arxiv.org/abs/1706.05853}{{\tt arXiv:1706.05853}}.
%Type = Article
\bibitem[{Hatta et~al.(2018)Hatta, Rajan, and Tanaka}]{Hatta:2018sqd}
\bibinfo{author}{Y.~Hatta}, \bibinfo{author}{A.~Rajan},
  \bibinfo{author}{K.~Tanaka},
\newblock \bibinfo{title}{{Quark and gluon contributions to the QCD trace
  anomaly}},
\newblock \bibinfo{journal}{JHEP} \bibinfo{volume}{12} (\bibinfo{year}{2018})
  \bibinfo{pages}{008}. \DOIprefix\doi{10.1007/JHEP12(2018)008}.
  \href{http://arxiv.org/abs/1810.05116}{{\tt arXiv:1810.05116}}.
%Type = Article
\bibitem[{Novikov and Shifman(1981)}]{Novikov:1980fa}
\bibinfo{author}{V.~A. Novikov}, \bibinfo{author}{M.~A. Shifman},
\newblock \bibinfo{title}{{Comment on the $\psi' \to J/\psi \pi \pi$ Decay}},
\newblock \bibinfo{journal}{Z. Phys. C} \bibinfo{volume}{8}
  (\bibinfo{year}{1981}) \bibinfo{pages}{43}.
  \DOIprefix\doi{10.1007/BF01429829}.
%Type = Article
\bibitem[{Donoghue and Leutwyler(1991)}]{Donoghue:1991qv}
\bibinfo{author}{J.~F. Donoghue}, \bibinfo{author}{H.~Leutwyler},
\newblock \bibinfo{title}{{Energy and momentum in chiral theories}},
\newblock \bibinfo{journal}{Z. Phys. C} \bibinfo{volume}{52}
  (\bibinfo{year}{1991}) \bibinfo{pages}{343--351}.
  \DOIprefix\doi{10.1007/BF01560453}.
%Type = Article
\bibitem[{Raman(1971)}]{Raman:1971jg}
\bibinfo{author}{K.~Raman},
\newblock \bibinfo{title}{{Gravitational form-factors of pseudoscalar mesons,
  stress-tensor-current commutation relations, and deviations from tensor- and
  scalar-meson dominance}},
\newblock \bibinfo{journal}{Phys. Rev. D} \bibinfo{volume}{4}
  (\bibinfo{year}{1971}) \bibinfo{pages}{476--488}.
  \DOIprefix\doi{10.1103/PhysRevD.4.476}.
%Type = Article
\bibitem[{Tong et~al.(2021)Tong, Ma, and Yuan}]{Tong:2021ctu}
\bibinfo{author}{X.-B. Tong}, \bibinfo{author}{J.-P. Ma},
  \bibinfo{author}{F.~Yuan},
\newblock \bibinfo{title}{{Gluon gravitational form factors at large momentum
  transfer}},
\newblock \bibinfo{journal}{Phys. Lett. B} \bibinfo{volume}{823}
  (\bibinfo{year}{2021}) \bibinfo{pages}{136751}.
  \DOIprefix\doi{10.1016/j.physletb.2021.136751}.
  \href{http://arxiv.org/abs/2101.02395}{{\tt arXiv:2101.02395}}.
%Type = Article
\bibitem[{Tong et~al.(2022)Tong, Ma, and Yuan}]{Tong:2022zax}
\bibinfo{author}{X.-B. Tong}, \bibinfo{author}{J.-P. Ma},
  \bibinfo{author}{F.~Yuan},
\newblock \bibinfo{title}{{Perturbative calculations of gravitational form
  factors at large momentum transfer}},
\newblock \bibinfo{journal}{JHEP} \bibinfo{volume}{10} (\bibinfo{year}{2022})
  \bibinfo{pages}{046}. \DOIprefix\doi{10.1007/JHEP10(2022)046}.
  \href{http://arxiv.org/abs/2203.13493}{{\tt arXiv:2203.13493}}.
%Type = Article
\bibitem[{Lepage and Brodsky(1980)}]{Lepage:1980fj}
\bibinfo{author}{G.~P. Lepage}, \bibinfo{author}{S.~J. Brodsky},
\newblock \bibinfo{title}{{Exclusive Processes in Perturbative Quantum
  Chromodynamics}},
\newblock \bibinfo{journal}{Phys. Rev. D} \bibinfo{volume}{22}
  (\bibinfo{year}{1980}) \bibinfo{pages}{2157}.
  \DOIprefix\doi{10.1103/PhysRevD.22.2157}.
%Type = Article
\bibitem[{Donoghue et~al.(1990)Donoghue, Gasser, and
  Leutwyler}]{Donoghue:1990xh}
\bibinfo{author}{J.~F. Donoghue}, \bibinfo{author}{J.~Gasser},
  \bibinfo{author}{H.~Leutwyler},
\newblock \bibinfo{title}{{The Decay of a Light Higgs Boson}},
\newblock \bibinfo{journal}{Nucl. Phys. B} \bibinfo{volume}{343}
  (\bibinfo{year}{1990}) \bibinfo{pages}{341--368}.
  \DOIprefix\doi{10.1016/0550-3213(90)90474-R}.
%Type = Article
\bibitem[{Nieves and Ruiz~Arriola(2000)}]{Nieves:1999bx}
\bibinfo{author}{J.~Nieves}, \bibinfo{author}{E.~Ruiz~Arriola},
\newblock \bibinfo{title}{{Bethe-Salpeter approach for unitarized chiral
  perturbation theory}},
\newblock \bibinfo{journal}{Nucl. Phys. A} \bibinfo{volume}{679}
  (\bibinfo{year}{2000}) \bibinfo{pages}{57--117}.
  \DOIprefix\doi{10.1016/S0375-9474(00)00321-3}.
  \href{http://arxiv.org/abs/hep-ph/9907469}{{\tt arXiv:hep-ph/9907469}}.
%Type = Article
\bibitem[{Nieves and {Ruiz Arriola}(2009)}]{Nieves:2009kh}
\bibinfo{author}{J.~Nieves}, \bibinfo{author}{E.~{Ruiz Arriola}},
\newblock \bibinfo{title}{{Meson Resonances at large $N_c$: Complex Poles vs
  Breit-Wigner Masses}},
\newblock \bibinfo{journal}{Phys. Lett.} \bibinfo{volume}{B679}
  (\bibinfo{year}{2009}) \bibinfo{pages}{449--453}.
  \DOIprefix\doi{10.1016/j.physletb.2009.08.021}.
  \href{http://arxiv.org/abs/0904.4590 [hep-ph]}{{\tt arXiv:0904.4590
  [hep-ph]}}.
%Type = Article
\bibitem[{Ecker et~al.(1989{\natexlab{a}})Ecker, Gasser, Leutwyler, Pich, and
  de~Rafael}]{Ecker:1989yg}
\bibinfo{author}{G.~Ecker}, \bibinfo{author}{J.~Gasser},
  \bibinfo{author}{H.~Leutwyler}, \bibinfo{author}{A.~Pich},
  \bibinfo{author}{E.~de~Rafael},
\newblock \bibinfo{title}{Chiral lagrangians for massive spin 1 fields},
\newblock \bibinfo{journal}{Phys. Lett.} \bibinfo{volume}{B223}
  (\bibinfo{year}{1989}{\natexlab{a}}) \bibinfo{pages}{425}.
%Type = Article
\bibitem[{Ecker et~al.(1989{\natexlab{b}})Ecker, Gasser, Pich, and
  de~Rafael}]{Ecker:1988te}
\bibinfo{author}{G.~Ecker}, \bibinfo{author}{J.~Gasser},
  \bibinfo{author}{A.~Pich}, \bibinfo{author}{E.~de~Rafael},
\newblock \bibinfo{title}{The role of resonances in chiral perturbation
  theory},
\newblock \bibinfo{journal}{Nucl. Phys.} \bibinfo{volume}{B321}
  (\bibinfo{year}{1989}{\natexlab{b}}) \bibinfo{pages}{311}.
%Type = Inproceedings
\bibitem[{Pich(2002)}]{Pich:2002xy}
\bibinfo{author}{A.~Pich},
\newblock \bibinfo{title}{{Colorless mesons in a polychromatic world}},
\newblock in: \bibinfo{booktitle}{{The Phenomenology of Large $N_c$ QCD}},
  \bibinfo{year}{2002}, pp. \bibinfo{pages}{239--258}.
  \DOIprefix\doi{10.1142/9789812776914\_0023}.
  \href{http://arxiv.org/abs/hep-ph/0205030}{{\tt arXiv:hep-ph/0205030}}.
%Type = Article
\bibitem[{Ledwig et~al.(2014)Ledwig, Nieves, Pich, Ruiz~Arriola, and Ruiz~de
  Elvira}]{Ledwig:2014cla}
\bibinfo{author}{T.~Ledwig}, \bibinfo{author}{J.~Nieves},
  \bibinfo{author}{A.~Pich}, \bibinfo{author}{E.~Ruiz~Arriola},
  \bibinfo{author}{J.~Ruiz~de Elvira},
\newblock \bibinfo{title}{{Large-$N_c$ naturalness in coupled-channel
  meson-meson scattering}},
\newblock \bibinfo{journal}{Phys. Rev. D} \bibinfo{volume}{90}
  (\bibinfo{year}{2014}) \bibinfo{pages}{114020}.
  \DOIprefix\doi{10.1103/PhysRevD.90.114020}.
  \href{http://arxiv.org/abs/1407.3750}{{\tt arXiv:1407.3750}}.
%Type = Article
\bibitem[{Toublan(1996)}]{Toublan:1995bk}
\bibinfo{author}{D.~Toublan},
\newblock \bibinfo{title}{{Lowest tensor meson resonances contributions to the
  chiral perturbation theory low-energy coupling constants}},
\newblock \bibinfo{journal}{Phys. Rev. D} \bibinfo{volume}{53}
  (\bibinfo{year}{1996}) \bibinfo{pages}{6602--6607}.
  \DOIprefix\doi{10.1103/PhysRevD.53.6602}.
  \href{http://arxiv.org/abs/hep-ph/9509217}{{\tt arXiv:hep-ph/9509217}},
  \bibinfo{note}{[Erratum: Phys.Rev.D 57, 4495 (1998)]}.
%Type = Article
\bibitem[{Ecker and Zauner(2007)}]{Ecker:2007us}
\bibinfo{author}{G.~Ecker}, \bibinfo{author}{C.~Zauner},
\newblock \bibinfo{title}{{Tensor meson exchange at low energies}},
\newblock \bibinfo{journal}{Eur. Phys. J. C} \bibinfo{volume}{52}
  (\bibinfo{year}{2007}) \bibinfo{pages}{315--323}.
  \DOIprefix\doi{10.1140/epjc/s10052-007-0372-x}.
  \href{http://arxiv.org/abs/0705.0624}{{\tt arXiv:0705.0624}}.
%Type = Article
\bibitem[{Nieves et~al.(2011)Nieves, Pich, and Ruiz~Arriola}]{Nieves:2011gb}
\bibinfo{author}{J.~Nieves}, \bibinfo{author}{A.~Pich},
  \bibinfo{author}{E.~Ruiz~Arriola},
\newblock \bibinfo{title}{{Large-$N_c$ Properties of the $\rho$ and $f_0(600)$
  Mesons from Unitary Resonance Chiral Dynamics}},
\newblock \bibinfo{journal}{Phys. Rev. D} \bibinfo{volume}{84}
  (\bibinfo{year}{2011}) \bibinfo{pages}{096002}.
  \DOIprefix\doi{10.1103/PhysRevD.84.096002}.
  \href{http://arxiv.org/abs/1107.3247}{{\tt arXiv:1107.3247}}.
%Type = Article
\bibitem[{Scadron(1968)}]{Scadron:1968zz}
\bibinfo{author}{M.~D. Scadron},
\newblock \bibinfo{title}{{Covariant Propagators and Vertex Functions for Any
  Spin}},
\newblock \bibinfo{journal}{Phys. Rev.} \bibinfo{volume}{165}
  (\bibinfo{year}{1968}) \bibinfo{pages}{1640--1647}.
  \DOIprefix\doi{10.1103/PhysRev.165.1640}.
%Type = Book
\bibitem[{Novozhilov(1975)}]{Novozhilov:1975yt}
\bibinfo{author}{Y.~V. Novozhilov}, \bibinfo{title}{{Introduction to Elementary
  Particle Theory}}, International Series of Monographs In Natural Philosophy,
  \bibinfo{publisher}{Pergamon Press}, \bibinfo{address}{Oxford, UK},
  \bibinfo{year}{1975}.
%Type = Article
\bibitem[{Moussallam(2011)}]{Moussallam:2011zg}
\bibinfo{author}{B.~Moussallam},
\newblock \bibinfo{title}{{Couplings of light I=0 scalar mesons to simple
  operators in the complex plane}},
\newblock \bibinfo{journal}{Eur. Phys. J. C} \bibinfo{volume}{71}
  (\bibinfo{year}{2011}) \bibinfo{pages}{1814}.
  \DOIprefix\doi{10.1140/epjc/s10052-011-1814-z}.
  \href{http://arxiv.org/abs/1110.6074}{{\tt arXiv:1110.6074}}.
%Type = Article
\bibitem[{Hoferichter et~al.(2024)Hoferichter, de~Elvira, Kubis, and
  Mei\ss{}ner}]{Hoferichter:2023mgy}
\bibinfo{author}{M.~Hoferichter}, \bibinfo{author}{J.~R. de~Elvira},
  \bibinfo{author}{B.~Kubis}, \bibinfo{author}{U.-G. Mei\ss{}ner},
\newblock \bibinfo{title}{{Nucleon resonance parameters from
  Roy\textendash{}Steiner equations}},
\newblock \bibinfo{journal}{Phys. Lett. B} \bibinfo{volume}{853}
  (\bibinfo{year}{2024}) \bibinfo{pages}{138698}.
  \DOIprefix\doi{10.1016/j.physletb.2024.138698}.
  \href{http://arxiv.org/abs/2312.15015}{{\tt arXiv:2312.15015}}.
%Type = Article
\bibitem[{Pham and Truong(1977)}]{Pham:1976yi}
\bibinfo{author}{T.~N. Pham}, \bibinfo{author}{T.~N. Truong},
\newblock \bibinfo{title}{{Muskhelishvili-Omnes Integral Equation with
  Inelastic Unitarity: Single and Coupled Channel Equations}},
\newblock \bibinfo{journal}{Phys. Rev. D} \bibinfo{volume}{16}
  (\bibinfo{year}{1977}) \bibinfo{pages}{896}.
  \DOIprefix\doi{10.1103/PhysRevD.16.896}.
%Type = Article
\bibitem[{Truong and Willey(1989)}]{Truong:1989my}
\bibinfo{author}{T.~N. Truong}, \bibinfo{author}{R.~S. Willey},
\newblock \bibinfo{title}{{Branching Ratios for Decays of Light Higgs Bosons}},
\newblock \bibinfo{journal}{Phys. Rev. D} \bibinfo{volume}{40}
  (\bibinfo{year}{1989}) \bibinfo{pages}{3635}.
  \DOIprefix\doi{10.1103/PhysRevD.40.3635}.
%Type = Article
\bibitem[{Celis et~al.(2014)Celis, Cirigliano, and Passemar}]{Celis:2013xja}
\bibinfo{author}{A.~Celis}, \bibinfo{author}{V.~Cirigliano},
  \bibinfo{author}{E.~Passemar},
\newblock \bibinfo{title}{{Lepton flavor violation in the Higgs sector and the
  role of hadronic $\tau$-lepton decays}},
\newblock \bibinfo{journal}{Phys. Rev. D} \bibinfo{volume}{89}
  (\bibinfo{year}{2014}) \bibinfo{pages}{013008}.
  \DOIprefix\doi{10.1103/PhysRevD.89.013008}.
  \href{http://arxiv.org/abs/1309.3564}{{\tt arXiv:1309.3564}}.
%Type = Article
\bibitem[{Becker et~al.(1979)Becker, Blanar, Blum, Cerrada, Dietl, Gallivan,
  Gottschalk, Grayer, Hentschel, Lorenz et~al.}]{becker1979model}
\bibinfo{author}{H.~Becker}, \bibinfo{author}{G.~Blanar},
  \bibinfo{author}{W.~Blum}, \bibinfo{author}{M.~Cerrada},
  \bibinfo{author}{H.~Dietl}, \bibinfo{author}{J.~Gallivan},
  \bibinfo{author}{B.~Gottschalk}, \bibinfo{author}{G.~Grayer},
  \bibinfo{author}{G.~Hentschel}, \bibinfo{author}{E.~Lorenz}, et~al.,
\newblock \bibinfo{title}{{A model-independent partial-wave analysis of the
  $\pi+ \pi$-system produced at low four-momentum transfer in the reaction
  $\pi^- p \to \pi^+ \pi^- n$ at 17.2 GeV/c}},
\newblock \bibinfo{journal}{Nucl. Phys. B} \bibinfo{volume}{151}
  (\bibinfo{year}{1979}) \bibinfo{pages}{46--70}.
%Type = Article
\bibitem[{Ananthanarayan et~al.(2001)Ananthanarayan, Colangelo, Gasser, and
  Leutwyler}]{Ananthanarayan:2000ht}
\bibinfo{author}{B.~Ananthanarayan}, \bibinfo{author}{G.~Colangelo},
  \bibinfo{author}{J.~Gasser}, \bibinfo{author}{H.~Leutwyler},
\newblock \bibinfo{title}{{Roy equation analysis of pi pi scattering}},
\newblock \bibinfo{journal}{Phys. Rept.} \bibinfo{volume}{353}
  (\bibinfo{year}{2001}) \bibinfo{pages}{207--279}.
  \DOIprefix\doi{10.1016/S0370-1573(01)00009-6}.
  \href{http://arxiv.org/abs/hep-ph/0005297}{{\tt arXiv:hep-ph/0005297}}.
%Type = Article
\bibitem[{Garcia-Martin et~al.(2011)Garcia-Martin, Kaminski, Pelaez, Ruiz~de
  Elvira, and Yndurain}]{Garcia-Martin:2011iqs}
\bibinfo{author}{R.~Garcia-Martin}, \bibinfo{author}{R.~Kaminski},
  \bibinfo{author}{J.~R. Pelaez}, \bibinfo{author}{J.~Ruiz~de Elvira},
  \bibinfo{author}{F.~J. Yndurain},
\newblock \bibinfo{title}{{The Pion-pion scattering amplitude. IV: Improved
  analysis with once subtracted Roy-like equations up to 1100 MeV}},
\newblock \bibinfo{journal}{Phys. Rev. D} \bibinfo{volume}{83}
  (\bibinfo{year}{2011}) \bibinfo{pages}{074004}.
  \DOIprefix\doi{10.1103/PhysRevD.83.074004}.
  \href{http://arxiv.org/abs/1102.2183}{{\tt arXiv:1102.2183}}.
%Type = Article
\bibitem[{Hanhart et~al.(2008)Hanhart, Pelaez, and Rios}]{Hanhart:2008mx}
\bibinfo{author}{C.~Hanhart}, \bibinfo{author}{J.~R. Pelaez},
  \bibinfo{author}{G.~Rios},
\newblock \bibinfo{title}{{Quark mass dependence of the rho and sigma from
  dispersion relations and Chiral Perturbation Theory}},
\newblock \bibinfo{journal}{Phys. Rev. Lett.} \bibinfo{volume}{100}
  (\bibinfo{year}{2008}) \bibinfo{pages}{152001}.
  \DOIprefix\doi{10.1103/PhysRevLett.100.152001}.
  \href{http://arxiv.org/abs/0801.2871}{{\tt arXiv:0801.2871}}.
%Type = Article
\bibitem[{Briceno et~al.(2018)Briceno, Dudek, Edwards, and
  Wilson}]{Briceno:2017qmb}
\bibinfo{author}{R.~A. Briceno}, \bibinfo{author}{J.~J. Dudek},
  \bibinfo{author}{R.~G. Edwards}, \bibinfo{author}{D.~J. Wilson},
\newblock \bibinfo{title}{{Isoscalar $\pi\pi, K\overline{K}, \eta\eta$
  scattering and the $\sigma, f_0, f_2$ mesons from QCD}},
\newblock \bibinfo{journal}{Phys. Rev. D} \bibinfo{volume}{97}
  (\bibinfo{year}{2018}) \bibinfo{pages}{054513}.
  \DOIprefix\doi{10.1103/PhysRevD.97.054513}.
  \href{http://arxiv.org/abs/1708.06667}{{\tt arXiv:1708.06667}}.
%Type = Article
\bibitem[{Donoghue and Na(1997)}]{Donoghue:1996bt}
\bibinfo{author}{J.~F. Donoghue}, \bibinfo{author}{E.~S. Na},
\newblock \bibinfo{title}{{Asymptotic limits and structure of the pion
  form-factor}},
\newblock \bibinfo{journal}{Phys. Rev. D} \bibinfo{volume}{56}
  (\bibinfo{year}{1997}) \bibinfo{pages}{7073--7076}.
  \DOIprefix\doi{10.1103/PhysRevD.56.7073}.
  \href{http://arxiv.org/abs/hep-ph/9611418}{{\tt arXiv:hep-ph/9611418}}.
%Type = Article
\bibitem[{Ruiz~Arriola and Sanchez-Puertas(2024)}]{RuizArriola:2024gwb}
\bibinfo{author}{E.~Ruiz~Arriola}, \bibinfo{author}{P.~Sanchez-Puertas},
\newblock \bibinfo{title}{{Phase of the electromagnetic form factor of the
  pion}},
\newblock \bibinfo{journal}{Phys. Rev. D} \bibinfo{volume}{110}
  (\bibinfo{year}{2024}) \bibinfo{pages}{054003}.
  \DOIprefix\doi{10.1103/PhysRevD.110.054003}.
  \href{http://arxiv.org/abs/2403.07121}{{\tt arXiv:2403.07121}}.
%Type = Article
\bibitem[{Krutov and Troitsky(2023)}]{Krutov:2023ztx}
\bibinfo{author}{A.~F. Krutov}, \bibinfo{author}{V.~E. Troitsky},
\newblock \bibinfo{title}{{Pion gravitational form factors at large momentum
  transfer in the instant-form relativistic impulse approximation approach}},
\newblock \bibinfo{journal}{Phys. Rev. D} \bibinfo{volume}{108}
  (\bibinfo{year}{2023}) \bibinfo{pages}{094043}.
  \DOIprefix\doi{10.1103/PhysRevD.108.094043}.
  \href{http://arxiv.org/abs/2310.14287}{{\tt arXiv:2310.14287}}.

\end{thebibliography}

\end{document}